\documentstyle[12pt,iopconf]{article}
\begin{document}
\begin{flushright}
CERN-TH/2002-374 \\
hep-ph/0212088
\end{flushright}

\title{Primordial Magnetic Fields \\
and Stochastic GW Backgrounds}

\author{Massimo Giovannini
\dag\footnote{E-mail : massimo.giovannini@cern.ch}, }

\affil{\dag\ Theoretical Physics Division, CERN, CH-1211, Geneva 23, 
Switzerland} 

\beginabstract
Operating resonant mass detectors 
set interesting bounds on diffused backgrounds of 
gravitational radiation and in the next five years the 
wide-band interferometers will also look for stochastic sources.
In this lecture the interplay among relic GW backgrounds 
and large scale magnetic fields will be discussed. 
Magnetic fields may significantly affect the thermal history
of the Universe in particular at the epoch electroweak 
symmetry breaking and shortly after. A review of  some old an new results 
on the spectral properties of stochastic GW backgrounds will be presented. 
The possible r\^ole of primordial magnetic fields as a source 
of gravitational radiation will be outlined. It will 
be shown that the usual bound on stochastic GW backgrounds 
 coming from the standard big bang 
nucleosynthesis (BBN) scenario can be significantly relaxed.
\endabstract
\section{Introduction}
The aim of this lecture is to outline the possible interplay
between primordial magnetic fields and relic backgrounds 
of gravitational radiation.
In section 2 the essential features of our magnetized local Universe will 
be outlined with the aim of supporting the idea that some primordial 
magnetic field had to exist prior to the decoupling of the radiation 
from matter.
 In section 3 the main spectral properties of stochastic GW backgrounds 
will be reviewed with special emphasis on the present goals 
of direct experimental search and on the  foreseen primordial signals.
In section 4 it will be argued that primordial 
magnetic fields can have an (indirect) impact on the formation of light nuclei.
This observation leads to a scenario of BBN which is interesting in its 
own right: the BBN with matter--antimatter domains. 
Section 5 contains our concluding remarks.

\section{A magnetized local Universe}

The first 
speculations concerning large scale magnetic fields are contained in a
seminal paper of Fermi \cite{f1} whose idea was that cosmic rays 
are in equilibrium not with the sun (or more generically with stars), 
as proposed by Alfv\'en \cite{f2}, 
but with the whole galaxy. In \cite{f1} the Milky Way was viewed as a 
magnetized gravitationally bound system with a large scale field  
${\cal O}(\mu {\rm G})$. Later this model was further 
explored  in collaboration 
with S. Chandrasekar \cite{f3} trying to connect galactic magnetic field 
an galactic angular momentum. Today, more than half a century after these 
pioneering attempts we know that some of the ideas are still valid.
For a more complete (but still too short) review 
on the subject of large scale magnetic fields in cosmology we refer the reader to some 
recent review written by the author \cite{g1} and to some excellent 
reviews \cite{rev3,rev4,rev5,rev6,rev7} concerning the astronomical sides of this 
manifold problem. In the following only few important points 
will be discussed.

Astronomical magnetic fields are certainly strong enough to influence the 
dynamics of gas in the galaxies and could have  been important during the 
formation of the galaxy. We do know that  galaxies are not the only magnetized
gravitationally bound systems in the Universe.
Clusters of galaxies, 
the intra-cluster medium and  the local super-cluster seems to be all 
magnetized.
The present knowledge is a result of the progress of fifty years 
of amazing achievements
 in astronomy. The first attempts of measuring magnetic fields 
came through optical polarization of starlight assuming that magnetic 
fields are aligned along the dust grains \cite{dav}.
The development of radio-astronomy 
made possible accurate determinations of magnetic fields in the 
interstellar medium 
through Faraday Rotation measurements \cite{rev3,rev6}. 
Polarization of synchrotron emission
allowed us to estimate the random component of large scale fields \cite{rev7}.
Observations of x-ray satellites produced  more reliable ``maps''
 of relativistic electrons in the intra-cluster medium leading, ultimately,
to a rather compelling evidence of a magnetized  medium on associated with 
individual galaxies 
\cite{cl5}.

\subsection{A controversial origin}

In spite of the amazing experimental achievements, the origin 
of large-scale magnetic fields remains still controversial and observational 
tests that could discriminate between competing theories represent a challenge 
for existing astronomical facilities. 
In particular, it seems puzzling that over very different length scales, from 
galaxies to super-clusters,  the magnetic field strength is always 
 ${\cal O}(\mu {\rm G})$. By itself the $\mu$ G field strength may indicate 
that magnetic fields are, today, all in {\em equipartition}, i.e. the idea that 
magnetic and kinetic energy densities may be, after all, comparable.  
Today, in fact 
roughly $B^2 \simeq T_{\rm cmb}^4$, where $T_{\rm cmb}$ is the Cosmic 
Microwave Background (CMB) temperature.  
The relativistic electron density is sometimes 
estimated using  equipartition. 
However, equipartition is not an experimental
evidence, it is a working hypothesis which may or may not be 
realized in the system under observation. For instance equipartition 
probably holds for the Milky Way but it does not seem to be valid in 
the Magellanic Clouds \cite{magcl}. The average equipartition 
field strengths in galaxies ranges from the $4 \mu$ G of M33 up to the 
$19 \mu$ G of NGC2276 \cite{eq}.

As far as galactic magnetic fields are concerned the common lore 
is that initially small magnetic inhomogeneities (with large 
correlation scale, of the order of the fraction of the Mpc) 
are amplified thanks to the global rotation of the galaxy. This 
is, {\em in nuce},  the dynamo theory. The dynamo theory 
has many important aspects (see \cite{kul} for a critical review) and it is 
formulated in the framework of non-relativistic magnetohydrodynamics
(MHD). Thanks to the linearity of MHD equations in the 
mean magnetic fields, some initial conditions for the galactic magnetic field 
{\em must} be postulated. This is the so-called initial {\em seed} hypothesis.
Furthermore, by roughly comparing the rotation period of the galaxy 
with its age we are led to conclude that the maximal achievable 
amplification one can obtain is of the order of the exponential
of the number of rotations performed by the galaxy, i.e. ${\cal O}(30)$. During the 
gravitational collapse of the protogalaxy, the magnetic flux is approximately 
conserved. Since the physical size of the system shrinks during 
collapse, then the magnetic field increases.

If this picture would be true, 
 the presently observed magnetic field in galaxies could 
be the result of the amplification of an initial seed as small as
${\cal O}(10^{-23} {\rm G})$, coherent over the scale of the gravitational collapse 
of the protogalaxy. However, numbers have to be much less generous. 
We assumed {\em exact} flux freezing during gravitational collapse. This 
is not always the case. Furthermore, it has been realized that 
thanks to the dynamo action not only large-scale fields are 
amplified but also small scale fields increase substantially their 
amplitude. Eventually the small scale fields may swamp the large-scale dynamo action.
 According to recent estimates $B_{\rm seed} \simeq {\cal O} (10^{-18} {\rm G})$.
While the beginning of the dynamo is fixed, essentially, by initial 
conditions, its end is due  to back-reaction effects driving 
the large-scale magnetic field on its {\em approximate} 
equipartition value. 

Even if the dynamo would be the correct explanation for the magnetic field 
of the galaxy, it is still a puzzle why
 clusters should have magnetic fields of the same strength. Clusters 
rotate much less than galaxies and dynamo theory would not give a significant 
increase in the amplitude of a primeval large scale field.
Furthermore,  experimental evidence tells us that there are magnetic 
fields in clusters which are not associated with individual galaxies. 
One can certainly think of some magnetic reconnection mechanism, 
analogous to the one of flares where the sun ejects magnetic flux 
together with plasma. In this case galaxies would eject magnetic flux in the 
intra-cluster medium. This idea does not tell why magnetic fields 
in the intra-cluster medium are coherent over $100$ kpc.

These rather peculiar features of our magnetized Universe
seem to indicate that large-scale magnetic fields may have a 
 primordial nature.
Even assuming that cluster and  supercluster magnetic fields are 
generated by some different astrophysical mechanism which 
by chance leads to a ${\cal O}(\mu {\rm G})$ field,
specific initial conditions for the dynamo mechanism should be 
anyway postulated in order to explain the origin of 
the galactic magnetic field. It would seem 
rather easy to get a magnetic field ${\cal O}(10^{-18} {\rm G})$ at the 
onset of gravitational collapse. This is not the case. In fact, even if 
the field has, apparently,  small amplitude (if compared, for instance, 
to the terrestrial magnetic field which is ${\cal O}({\rm G})$) 
its correlation scale must be huge. Notice 
 that a quantum mechanical fluctuation
in a box of 1 Mpc is ${\cal O}(10^{-60}{\rm G})$..

Different possibilities can then be envisaged.
\begin{itemize}
\item Large-scale magnetic fields are {\em not} primordial and are only 
explained on astrophysical basis. This possibility
requires initial conditions for the dynamo evolution. 
These initial conditions have to be postulated.
\item There was some (small) primordial field resulting in a 
${\cal O}(10^{-18} {\rm G})$ at the time of gravitational collapse 
over the typical scale of the collapse. Then galactic magnetic fields 
may be generated via dynamo. 
\item Large-scale magnetic fields are the result of a primordial
field ${\cal O}(10^{-10} {\rm G})$ present at the onset of 
gravitational collapse. In this case the dynamo action would not 
even be required and the problem of magnetic fields in clusters
will be relaxed, if not completely solved.
\end{itemize}
Among all these possibilities it is very hard to decide on the basis 
of {\em simulations} or on the basis of {\em theoretical constraints}.
A more clever way of discriminating among these ideas is 
to look at the sky and check if the predictions of the primordial theory 
are verified in nature. This is difficult but not impossible since 
the fully primordial theory (without dynamo action) predicts a specific 
parity of the field with respect to rotations by $\pi$ about the galactic center
which is different from the prediction of the fully primordial hypothesis.

\subsection{The primordial hypothesis}

The physical reason why magnetic fields, unlike other relics, are 
preserved by the cosmological evolution is 
that the Universe was (and partially still is) an extremely good conductor.
Consider, for simplicity, a (conformally flat) Friedmann-Robertson-Walker Universe 
\begin{equation}
ds^2 = a^2(\eta) [ d\eta^2 - d \vec{x}^2],
\label{m1}
\end{equation}
where $\eta$ is the conformal time coordinate and $a(\eta)$ is the scale factor 
which, for instance, evolves linearly during a radiation-dominated phase of expansion,
quadratically  during matter-domination, hyperbolically during and exact 
de Sitter phase.

In FRW space-time we can write, for instance, two of the basic laws 
of resistive MHD, also called, sometimes, Alfv\'en theorems
\begin{eqnarray}
&&\frac{d}{d\eta} \int_{\Sigma} \vec{B} \cdot d\vec{\Sigma}=-
\frac{1}{\sigma} \int_{\Sigma} \vec{\nabla} \times\vec{\nabla}
\times\vec{B}\cdot d\vec{\Sigma},
\label{flux}\\
&&\frac{d}{d\eta} {\cal H}_{M} = - \frac{1}{\sigma} \int_{V} d^3 x
{}~\vec{B}\cdot\vec{\nabla} \times\vec{B},
\label{hel}
\end{eqnarray}
where $\Sigma$ is an arbitrary closed surface which moves with the
plasma and where 
\begin{equation}
{\cal H}_{M} = \int_{V} d^3 x \vec{A}~\cdot \vec{B},
\label{h1}
\end{equation}
is the magnetic helicity and the integration volume $V$ is defined in such a way that
$\vec{B}$ is parallel to the surface $\partial V$ which bounds $V$. 
In Eqs. (\ref{hel})--(\ref{h1}) $\vec{A}$ is the vector potential
\footnote{ The quantities appearing in Eqs. (\ref{flux})--(\ref{h1}) 
are defined in curved space and they are related to their flat-space counterpart as :
$\vec{B} = a^2 \vec{{\cal B}}$, $\vec{A} = a \vec{{\cal A}}$, $\sigma = \sigma_{c}a$.} and 
$\sigma$ is the conductivity.

In the hot big bang model the temperature increases when we go back 
in time and so does the conductivity. In the limit of infinite 
conductivity (sometimes called ideal or superconducting 
limit) 
the right hand side of Eqs. (\ref{flux}) and (\ref{hel})
goes to zero. In the ideal limit both the magnetic flux and the magnetic 
helicity are {\em exactly} conserved and   Eq. (\ref{flux}) implies, 
 that the magnetic flux lines are always glued together with the 
plasma element, or, for short, that magnetic flux 
is frozen into the plasma element. From 
 Eq. (\ref{hel}), also the magnetic helicity is conserved if $\sigma \to \infty$.
This means  the number of knots and twists 
in the magnetic flux lines stays always the same. 
From the physical point of view, the two Alfv\'en theorems can be understood in
simple terms. 
Magnetic field lines must be closed (because of transversality). However 
there could be different topological situations.
For instance closed loops may have no intersections. In this case 
the helicity is zero. There could be however the situation where a loop 
is twisted (like some type of Moebius stripe) or the case where two loops are 
connected (like the rings of a chain). In these situations  
the magnetic helicity is non zero and it is conserved in the superconducting limit.
While the conservation of magnetic flux tell us that the {\em energetical} 
properties of the magnetic field distribution are conserved, the conservation 
of the magnetic helicity implies the conservation of the 
{\em topological} properties of the magnetic flux lines.

In the early Universe the contribution of the resistivity, i.e. $1/\sigma$ 
is never zero and, therefore, the conservation of the flux and of the 
helicity can be only  approximate. The quantity
\begin{equation}
r_{\rm B}(L) = \frac{\rho_{\rm B}(L,\eta)}{\rho_{\gamma}(\eta)} \simeq
 \frac{\langle |\vec{{\cal B}}(L, \eta)|^2 \rangle}{T^4(\eta)}, 
\label{R}
\end{equation}
is approximately conserved all along the time evolution since, because 
of of flux freezing $|{\vec {\cal B}}| \sim a^{-2}$ and, 
because of adiabatic evolution $T \sim a^{-1}$. In Eq. (\ref{R}) 
$L$ is the typical coherence scale of the field. 
In terms of $r_{\rm B}$ we can easily write 
the requirements discussed above in the context of the dynamo mechanism.
In particular, if we assume that the amplification of the magnetic field due to dynamo 
was ${\cal O}(30)$ e-folds and that flux was exactly frozen during the collapse 
of the protogalaxy, we are led to demand, in order to turn on successfully the dynamo action, 
that $r_{\rm  B}\geq {\cal O}(10^{-34})$. As previously pointed out, these 
considerations are rather naive and the realistic requirement
is that $r_{\rm  B}\geq {\cal O}(10^{-24})$ corresponding to a field 
${\cal O}( 10^{-18} {\rm G})$ at the onset of gravitational collapse and 
over a typical scale $L \sim 1$ Mpc.

If the dynamo mechanism is invoked, the primordial content of the magnetic 
field sets the initial conditions of the MHD evolution. It could also 
happen that, since the generated magnetic fields are rather large, there is no 
need of dynamo amplification and all the amplification occurs during the 
gravitational collapse. Atypical value of primordial field may be of the order of 
$r_{\rm B} \sim 10^{-8}$--$10^{-9}$. 

Back in the late sixties Harrison
\cite{har} suggested that the initial conditions 
of the MHD  equations might have something 
to do with cosmology in the same way as it was
  suggested that the primordial spectrum of gravitational potential 
fluctuations (i.e. the Harrison-Zeldovich spectrum) 
might be produced in some primordial phase of the evolution of the Universe.
Since then, several mechanisms have been invoked 
in order to explain the origin of the magnetic seeds and few of them 
are compatible with inflationary evolution. It is not my purpose 
to review here all the different mechanisms which have been proposed so far
 (see, for 
instance, \cite{g1}).
The cosmological mechanisms can be
 {\em causal} mechanisms (if the magnetic seeds are produced at a given time inside the 
horizon) and {\em inflationary}  (if  correlations in the magnetic field 
are produced outside the horizon). Both classes 
of models have their own virtues and their own problems. 
Causal mechanisms, for instance, lead to large magnetic fields but over small 
length-scales and, typically, the scale of the relevant domains at the onset of gravitational collapse 
is much smaller than the Mpc. On the other hand, inflationary mechanisms 
can efficiently produce large magnetized domains but with 
very small field intensity.

From the physical point of view it would be desirable to have a model
where  
small quantum mechanical fluctuations of gauge fields are 
amplified thanks to the dynamical evolution. Then the amplified quantum 
mechanical fluctuations will become, eventually, the initial 
conditions of the MHD evolution.
This is in full analogy with what it is done with scalar fluctuations 
of the metric in the context 
of ordinary inflationary models. The current explanation of the detected CMB anisotropies
is indeed that they are the result of amplified fluctuations of the metric.
The problem with gauge fields is that their evolution equations 
are qualitatively very different from the evolution equations of the fluctuations of the 
metric. Quantum mechanical fluctuations 
of gauge fields in four space-time dimensions they are not  likely to be amplified.
This is one of the main motivations, in this context, in order 
to go beyond four dimensions and study 
if and how magnetic fields are generated 
when the gauge coupling is effectively time dependent \cite{g2,g3,g4,g5}.
In the context of pre-big bang models \cite{ven} $r_{\rm B}$ can be as large as 
$10^{-8}$ \cite{g2,g3}. If the variation of the gauge couplings
 occurs during a de Sitter stage 
of expansion we can get $r_{\rm B} \sim 10^{-12}$ \cite{g4}. 

In conclusion we can say that astrophysical 
mechanisms for the origin of large-scale magnetic fields 
have to rely on some initial conditions at the epoch of gravitational collapse.
Cosmology is able to provide these initial conditions 
in a number of different models. The situation is, in this sense, not 
different from what happens in the physics of CMB anisotropies. 
The common feature of various mechanisms producing magnetic fields 
in the early Universe is that magnetic fields are generated not only 
over the typical scale of the gravitational collapse 
of the protogalaxy but also over different physical scales. 
This means that the magnetic fields produced in the far past 
and giving, for instance, initial conditions for the 
dynamo mechanism, will also influence other moments of the 
thermodynamical history of the Universe. Two examples are 
 the moment of electroweak symmetry breaking and the moment of BBN. 

\subsection{Hypermagnetic knots and electroweak symmetry breaking}

Since  a generic magnetic field configuration at finite conductivity 
leads to an 
energy-momentum tensor which is anisotropic and which has non-vanishing 
transverse and traceless component (TT), 
if magnetic fields are present inside the horizon at some epoch they can 
radiate 
gravitationally. In more formal terms this statement can be understood since
the TT components of the energy momentum tensor acts as a source term for the 
TT fluctuations of the geometry which are associated with 
gravitational waves. A non-trivial example of this effect is 
provided by magnetic knot configurations \cite{gio7} which are 
transverse (magnetic) field configurations with 
a topologically non-trivial structure in the flux lines.

For sufficiently high temperatures and for sufficiently 
large values of the various 
fermionic charges the $SU(2)_{L}\otimes U(1)_{Y}$ symmetry is restored 
and 
non-screened vector modes will now correspond to the hypercharge group. 
Topologically non-trivial configurations of the hypermagnetic field 
($\vec{{\cal H}}_{Y}$)
 can be related to the 
baryon asymmetry of the Universe (BAU) \cite{shap,gio9,gio10,gio11} and they 
can also radiate 
gravitationally \cite{gio8,rub}. In this context the 
value of the BAU is directly related to the amplitude 
of the stochastic GW background. The evolution 
equations of the hypercharge field at finite conductivity 
imply that the largest modes which can survive in the plasma 
are the ones associated with the hypermagnetic conductivity 
frequency which is roughly eight orders of magnitude smaller 
than the temperature at the time of the electroweak 
phase transition which I take to occur  
around $100$ GeV.

If a hypermagnetic background is present for $T> T_c$, then
  the energy momentum tensor 
will acquire a small anisotropic component which will source the evolution 
equation of the tensor fluctuations $h_{\mu\nu}$ of the metric $g_{\mu\nu}$: 
\begin{equation}
h_{ij}'' + 2 {\cal H} h_{ij}' - \nabla^2 h_{ij} = - 16 \pi G
\tau^{(T)}_{ij}.
\label{GWeq}
\end{equation}
where $\tau^{(T)}_{ij}$ is the {\em tensor} component of the 
{\em energy-momentum tensor} 
of the hypermagnetic fields. Suppose now that 
$|\vec{{\cal H}}|$ has constant amplitude and that it is also 
homogeneous. Then  we can easily deduce 
the critical fraction of energy density  present today in relic gravitons 
of EW origin 
\begin{equation}
\Omega_{\rm gw}(t_0) = \frac{\rho_{\rm gw}}{\rho_c} 
\simeq z^{-1}_{{\rm eq}}
r^2,~~\rho_{c}(T_{c})\simeq N_{\rm eff} T^4_{c }
\end{equation}
($z_{\rm eq}$ is the redshift from the time of matter-radiation,
 equality and$ N_{\rm eff} = 106.75$ is the effective number of spin degrees of freedom at 
  $T_{c} \sim 100 $ GeV). Because of the structure of the equations describing the evolution 
of the system at finite fermionic density and finite conductivity \cite{gio10,gio11}, 
stable 
hypermagnetic fields will be present not only for 
$\omega_{\rm ew}\sim k_{\rm ew}/a$ but 
for all the range $\omega_{{\rm ew}} <\omega< \omega_{\sigma}$ where 
$\omega_{\sigma}$ is the diffusivity frequency.  
The (present) values of 
$\omega_{\rm ew}$ is 
\begin{equation}
\omega_{\rm ew } (t_0) \simeq 2.01 \times 10^{-7} \biggl( \frac{T_{c}}{1 {\rm GeV}} \biggr) 
\biggl(\frac{ N_{\rm eff}}{100} \biggr)^{1/6} {\rm Hz} .
\end{equation}
Thus, $\omega_{\sigma}(t_0) \sim 10^{8} \omega_{\rm ew} $. Suppose now that 
$T_{c} \sim 100$ GeV; than we will have that $\omega_{\rm ew}(t_0) \sim 10^{-5}$ Hz. 
Suppose now  that 
\begin{equation}
|\vec{{\cal H}}|/T_{c}^2 \geq 0.3.
\end{equation} 
This requirement imposes $ r \simeq 0.1$--$0.001$ and, consequently, 
\begin{equation}
h_0^2 \Omega_{\rm GW} \simeq 10^{-7} - 10^{-8}.
\end{equation}
Notice that this signal would occurr in a (present) frequency 
range between $10^{-5}$ and $10^{3}$ Hz. This signal 
satisfies the presently available phenomenological 
bounds on the graviton backgrounds of primordial origin (see the 
following section).
The pulsar timing bound  is automatically satisfied
since our hypermagnetic background is defined for $10^{-5} {\rm Hz} 
\leq \omega \leq 10^{3} {\rm Hz}$. The large-scale bounds would imply 
$h_0^2 \Omega_{\rm GW} < 7 \times 10^{-11}$ but a at much lower frequency 
(i.e. $10^{-18 }$ Hz). The signal discussed here is completely 
absent for frequencies $\omega < \omega_{\rm ew}$. Notice that 
this signal is clearly distinguishable from other stochastic 
backgrounds occurring at much higher frequencies (GHz region) 
like the ones predicted by quintessential inflation and pre-big bang 
cosmology (see following Section).
The frequency of operation of the interferometric devices 
(VIRGO/LIGO) is located between few Hz and 10 kHz.
 The frequency of operation 
of LISA is well below the Hz (i.e. $10^{-3} $Hz, approximately). 
 In this model the signal 
can be located both in the LISA window and in the VIRGO/LIGO window
due to the hierarchy between the hypermagnetic diffusivity scale and the 
horizon scale at the phase transition \cite{gio9,gio11}.

\section{Spectral Properties of stochastic GW backgrounds}

The fraction of critical energy 
density $\rho_c$ stored in relic gravitons at 
the present (conformal) time $\eta_0$ 
per each logarithmic interval of the physical frequency $f$
\begin{equation}
\Omega_{{\rm GW}}(f,\eta_0)\,=\,\frac{1}{\rho_{c}}\,
\frac{{\rm d} \rho_{{\rm GW}}}{{\rm d} \ln{f}}\,=\, 
\overline{\Omega}(\eta_0)\,\omega(f,\eta_0)\,
\label{Omegath}
\end{equation}
is the quantity we will be mostly interested in.

The frequency dependence in $\Omega_{\rm GW}(f,\eta_0)$ is a 
specific feature of the mechanism responsible 
for the production of the gravitons and, in
 a given interval of the present frequency, the slope of the 
logarithmic energy spectrum can be defined as 
\begin{equation}
\alpha = \frac{ d\, \ln{\omega(f,\eta_0)}}{d\, \ln{f}}.
\label{sl}
\end{equation}
If, in a given logarithmic interval of frequency, $\alpha <0 $ 
the spectrum is {\em red} since 
its dominant energetical content is stored in the infra-red. 
If, on the other hand $0<\alpha \leq 1$ the spectrum is 
{\em blue}, namely a mildly increasing logarithmic energy 
density. Finally if $ \alpha > 1$ we will talk about {\em violet} 
spectrum whose dominant energetical content is stored in the 
ultra-violet. The case $\alpha =0$ corresponds to the 
case of scale-invariant (Harrison-Zeldovich) logarithmic 
energy spectrum.

Every sudden variation of the background geometry from 
one regime of expansion to the other leads 
inevitably to the production of
graviton pairs which are stochastically distributed \cite{gr0,gr01,gr02}. 
Valuable reviews on the subject are given in Refs. 
\cite{revg1,revg2,revg3}.
The amplitude 
of the detectable signal depends, however, upon the 
specific model of curvature evolution. In ordinary 
inflationary models the amount of gravitons produced 
by a variation of the geometry is notoriously 
quite small. This feature can be traced back to the 
fact that $\Omega_{\rm GW}(f,\eta_0)$ is either a decreasing or 
(at most) a flat function of the present frequency. 
Suppose, for simplicity, that the ordinary inflationary phase is 
suddenly followed by a radiation dominated phase turning, after 
some time, into a matter dominated stage of expansion. The 
logarithmic energy spectrum will have, as a function of the present 
frequency, two main branches : an infra-red branch (roughly 
ranging between $10^{-18}$ Hz and $10^{-16}$ Hz) and a flat (or 
possibly decreasing) branch between $10^{-16}$ and $100$ MHz. 

The flat  branch of 
the spectrum is mainly due to those modes leaving the 
horizon during the inflationary phase and re-entering 
during the radiation dominated epoch. The infra-red branch of 
the spectrum is  produced by modes leaving the 
horizon during the inflationary phase and re-entering during the 
matter dominated epoch.

Starting from infra-red we have that the COBE observations of the 
first thirty multipole moments of the temperature 
fluctuations in the microwave sky imply that the GW contribution to the 
Sachs-Wolfe integral cannot exceed the amount of anisotropy 
directly detected. This implies that for frequencies $f_0$ 
approximately comparable with $H_0$ and 20 $H_0$ (where 
$H_0$ is the present value of the Hubble constant including 
its indetermination $h_0$) $ h_0^2 \, \Omega_{\rm GW} ( f_0 , \eta_0) 
< 7 \times 10^{-9} $.  
Moving toward the ultra-violet, the very small size of the 
fractional timing error in the arrivals of the 
millisecond pulsar's pulses requires that 
$\Omega_{\rm GW}(f_{P}, \eta_0) < 10^{-8}$ for a typical frequency 
roughly comparable with the inverse of the observation time 
during which the pulses have been monitored, i.e. $f_{P} \sim 10$ nHz.

Finally, if we believe the simplest (homogeneous and isotropic) big-bang
nucleosynthesis (BBN) scenario we have to require that the total fraction of 
critical energy density stored in relic gravitons at the BBN time 
does not exceed the energy density stored in relativistic matter at the 
same epoch. Defining $\Omega_{\gamma}(\eta_0)$ as the fraction of critical 
energy density presently stored in radiation we have that 
the BBN  consistency requirement demands 
\begin{equation} 
h^2_0\,\int^{f_{\rm max}}_{f_{\rm ns}}\,\Omega_{\rm GW}(f,\eta_0)\;{\rm d}\ln{f}\,
\leq\,5\,\times\,10^{-6} \,\Delta N_{\rm eff}
\label{bbn}
\end{equation}
where  $f_{\rm ns}\,\simeq\,0.1$ nHz is the present value 
of the frequency corresponding to the horizon at the nucleosynthesis 
time; $f_{\rm max}$ stands for  the maximal frequency of the spectrum 
and it depends upon the specific theoretical model (
in the case of ordinary inflationary models $f_{\rm max} = 100$ MHz).
In Eq. (\ref{bbn}), $\Delta N_{\rm eff}$ is the excess in the effective 
number of neutrino species which will be discussed in section 4.
The constraint expressed in Eq. (\ref{bbn}) is 
{\em global} in the sense that it bounds the {\em integral} of the 
logarithmic energy spectrum. The constraints coming from 
pulsar's timing errors and from the integrated Sachs-Wolfe effect
are instead {\em local} in the sense that they 
bound the value of the logarithmic energy spectrum in a specific interval of 
frequencies. 

In the case of stochastic GW backgrounds of 
 inflationary origin, owing to the red nature
of the logarithmic energy spectrum, the most significant constraints 
are the ones present in the soft region of the spectrum, 
more specifically, the 
ones connected with the Sachs-Wolfe effect. 
Taking into account the specific frequency behavior in the 
infra-red branch of the spectrum and assuming perfect scale invariance 
we have that $h_0^2\,\,\,\Omega_{\rm GW} ( f,\eta_0) < 10^{-15}$ for 
frequencies $f> 10^{-16}$ Hz. 
We have to conclude that 
the inflationary spectra are invisible by pairs of interferometric detectors 
operating in a 
window ranging approximately between few Hz and $10$ kHz. 

In order to illustrate more quantitatively this point we remind the 
expression of the signal-to-noise ratio (SNR)  in the 
context of optimal processing  required for the detection 
of stochastic backgrounds 
\cite{int1,int2,int3,int4}. By 
assuming that the intrinsic 
 noises of the detectors are stationary, Gaussian, 
uncorrelated, much larger in amplitude than the gravitational strain, and 
statistically independent on the strain itself, one has: 
\begin{equation}
{\rm SNR}^2 \,=\,\frac{3 H_0^2}{2 \sqrt{2}\,\pi^2}\,F\,\sqrt{T}\,
\left\{\,\int_0^{\infty}\,{\rm d} f\,\frac{\gamma^2 (f)\,\Omega^2_{{\rm GW}}(f)}
{f^6\,S_n^{\,(1)} (f)\,S_n^{\,(2)} (f)}\,\right\}^{1/2}\; ,
\label{2}
\end{equation}
($F$ depends upon 
the geometry of the two detectors and in the case of the correlation between 
two interferometers $F=2/5$; $T$ is the observation time). 
In Eq. (\ref{2}), $S_n^{\,(k)} (f)$ is the (one-sided) noise power 
spectrum (NPS) of the $k$-th 
$(k = 1,2)$ detector. The NPS contains the important informations 
concerning the 
noise sources (in broad terms seismic, thermal and shot noises)
 while $\gamma(f)$ is the overlap reduction function 
which is determined by the relative locations and orientations 
of the two detectors. Without going through the technical details 
\cite{gio3,dan2,dan3}
from the expression of the SNR we want to notice that the 
achievable sensitivity of a pair of wide band interferometers crucially 
depends upon the spectral slope of the theoretical energy spectrum in the 
operating window of the detectors. So, a flat spectrum will lead 
to an experimental sensitivity which might not be similar to the 
sensitivity achievable in the case of a blue or violet spectra 
\cite{gio3,gio4a,gio4b}. In the case of an exactly scale invariant spectrum
the correlation of the two (coaligned) LIGO detectors with 
central corner stations in Livingston (Lousiana) and in Hanford 
(Washington) will have a sensitivity to a flat spectrum 
which is $h_0^2\,\,\, \Omega_{\rm GW}(100~{\rm Hz}) 
\simeq 6.5 \times 10^{-11} $ 
after one year of observation and with signal-to-noise 
ratio equal to one \cite{gio3}. 
This implies that ordinary inflationary spectra 
are (and will be) invisible by wide band detectors since the 
inflationary prediction, in the most favorable case (i.e.
 scale invariant spectra), 
undershoots the experimental sensitivity by more than four orders of magnitude.

\subsection{Scaling violations in graviton spectra}

In order to have  a large detectable signal between $1$ Hz and $10$ kHz 
we have to 
look for models exhibiting scaling violations for frequencies larger than 
the mHz. The scaling violations should go in the direction 
of blue ($0< \alpha \leq 1$) or violet  ($\alpha > 1$) 
logarithmic energy spectra. Only in this case we shall have the hope that 
the signal will be large enough in the window 
of wide band detectors. Notice that the growth of the spectra should 
not necessarily be monotonic: we might have a blue or violet spectrum for 
a limited interval of frequencies with a spike or a hump.

Suppose now, as a toy example, that the ordinary inflationary 
phase is not immediately 
followed by a radiation dominated phase but by a quite long phase 
expanding slower than radiation \cite{gio1}. This speculation  is 
theoretically 
plausible since we ignore what was the thermodynamical history of the Universe 
prior to BBN. If the Universe expanded slower than radiation the equation of 
state 
of the effective sources driving the geometry had to be, for some time, 
stiffer than radiation. This means that the effective speed of sound $c_s$ 
had to lie in the range $1/\sqrt{3} < c_{s} \leq 1$.
Then the resulting logarithmic energy spectrum,  for the 
modes leaving the horizon during the inflationary phase and 
re-entering during the stiff phase, is tilted toward large 
frequencies with typical (blue) slope given by \cite{gio1}
\begin{equation}
\alpha = \frac{ 6 c_s^2 - 2 }{3 c_s^2 +1}\,,\,\,\,\,\, 0< \alpha \leq 1.
\end{equation}
A situation very similar to the one we just described occurs in 
quintessential inflationary models \cite{vil1}. In this case the 
tilt is maximal (i.e., $\alpha=1 $) and a more precise calculation shows 
the appearance of logarithmic corrections in the logarithmic 
energy spectrum which becomes \cite{gio4a,gio4b,gio1,vil1} 
$\omega(f) \propto f \ln^2{f}$.
The maximal frequency $f_{\rm max}(\eta_0)$ is of the order of $100$ GHz
(to be compared with the $100$ MHz of ordinary inflationary models)
and it corresponds to 
the typical frequency of a  spike in the GW background. In quintessential 
inflationary models the relic graviton background will then have 
the usual infra-red and flat branches supplemented, at high 
frequencies (larger than the mHz and smaller than the GHz) by a true 
hard branch \cite{gio4a,gio4b} whose peak can be, in terms of $h_0^2 \,\,\, 
\Omega_{\rm GW}$, 
 of the order of $10^{-6}$, compatible 
with the BBN bound and  roughly eight orders of magnitude larger than the 
signal provided by ordinary inflationary models. 

An interesting aspect of this class of models is that the maximal signal 
occurs in a frequency region between the MHz and the GHz. 
Microwave cavities can be used as GW detectors precisely in the 
mentioned frequency range \cite{cav1}. There were published results 
reporting the construction  of this type of detectors \cite{cav2} and the 
possibility of further improvements in the sensitivity received 
recently attention \cite{cav3}. Our signal is certainly a candidate 
for this type of devices.

\subsection{String cosmological models} 

In string cosmological models \cite{ven} of pre-big bang type 
$h_0^2 \,\,\,\Omega_{\rm GW}$ can 
be as large as $10^{-7}$--$10^{-6}$ for frequencies ranging between $1$ Hz and 
$100$ GHz \cite{gio2,gio2a,gio2b,gio2d}. 
In these types of models the logarithmic energy spectrum can be 
either blue or violet depending upon the given mode of the spectrum. If the mode
under consideration 
left the horizon during the dilaton-dominated epoch the typical 
slope will be violet (i.e. $ \alpha \sim 3 $ up to logarithmic corrections).
If the given mode left the horizon during the stringy phase the slope can 
be also blue with typical spectral slope $\alpha \sim 6 - 2 ( \ln{g_1/g_s}/\ln{z_s})$
where $g_1$ and $g_s$ are the values of the dilaton coupling at the 
end of the stringy phase and at the end of the dilaton dominated phase; $z_s$ 
parametrizes the duration of the stringy phase. This 
behaviour is representative of the minimal string cosmological 
scenarios. However, in the non-minimal case the spectra can also be non 
monotonic. Recently the sensitivity of a pair of VIRGO detectors
to string cosmological gravitons was specifically analyzed \cite{gio10} with the 
conclusion that a VIRGO pair, in its upgraded stage, will certainly be able to probe 
wide regions of the parameter space of these models. If we  maximize the 
overlap between the two detectors \cite{gio10} or 
if we would  reduce (selectively) the pendulum and pendulum's internal modes
contribution to the thermal noise of the instruments \cite{gio11}, the 
visible region (after one year of observation and with SNR equal to one)
of the parameter space will get even larger. Unfortunately, as in the 
case of the advanced LIGO detectors, also in the case of the advanced VIRGO 
detector the sensitivity to a flat spectrum will be irrelevant for 
ordinary inflationary models.

\section{Relaxing the BBN bound}

The strongest constraint on additional energy density in the universe
with a radiation-like equation of state
is provided by big bang nucleosynthesis (BBN).
The additional energy
density speeds up the expansion and cooling of the universe,
and, consequently,
the typical  time scale of BBN is reduced in comparison with the
standard case.
The additional
radiation-like energy density may be attributed to some extra relativistic
species whose statistics may be either bosonic or fermionic.
Since the supplementary species may be  fermionic,  they have been
customarily parametrized in terms of the effective number of
neutrino species
\begin{equation}
N_{\rm eff} = 3 + \Delta N_{\rm eff},
\end{equation}
where $\Delta N_{\rm eff} = 0$ corresponds to the standard case
with no extra energy density.
The standard BBN (SBBN) results are in agreement with the observed abundances
for $N_{\rm eff} = 2$--$4$, giving thus an upper limit $\Delta N_{\rm eff} \leq 1$.

If hypermagnetic fields are present at the electroweak time, matter--antimatter 
domains can be generated and persist until the time 
of BBN \cite{gio9,gio10}. This possibility is rather interesting since 
matter--antimatter domains suggest a slightly different 
scenario of BBN which has been developed independently on the motivation
stemming from hypermagnetic fields \cite{mam2,mam3,mam4,mam4a,mam5,hannu}.
 
Provided that matter--antimatter domains are present at the onset of big bang
nucleosynthesis (BBN), the number of allowed additional relativistic species
increases, compared to the standard scenario when matter--antimatter domains
are absent \cite{hannu}.
The extra relativistic species may take the form of massless
fermions or even massless bosons, like relic gravitons.
The number of additional degrees of
freedom compatible with BBN depends, in this framework,
  upon the typical scale of the domains and the
antimatter fraction. Since
the presence of matter--antimatter domains allows a reduction of the
neutron to proton ratio prior to the formation
of $^{4}{\rm He}$,  large amounts of radiation-like energy density
are allowed.
The present critical fraction of energy density stored in relic gravitons, i.e. 
(\ref{bbn})
depends upon $\Delta N_{\rm eff}$ whose range of variation
can  be translated
into constraints on the energy density of
relic gravitational waves produced prior to BBN.

Various resonant mass
detectors  are now operating
\cite{allegro,auriga,explorer,nautilus}.
In \cite{astone}, the first
experiment of cross-correlation between two cryogenic detectors
has been reported  with the purpose
of giving an upper limit on  $h_{0}^2 \Omega_{\rm GW}$. The two detectors are
Explorer \cite{explorer} (operating in CERN, Geneva) and Nautilus \cite{nautilus}
 (operating in Frascati,
near Rome). Previous experiments giving upper limits on $h_{0}^2 \Omega_{\rm GW}$
used room temperature detectors. The Rome group obtained then an
upper limit $h_{0}^2 \Omega_{\rm GW} < 60$ at a frequency of
roughly $905$ Hz. The limit is a result of cross-correlation between the two
detectors (located at a distance of approximately $600$ km) for an integration
time of approximately 12 hours. This limit is not competitive with the BBN bound
(and also  above the critical density bound implying that $\Omega_{\rm GW}<1$).
However, by increasing the correlation time from few hours
to few months it is not unreasonable to go below one in $ h_0^2\Omega_{\rm GW}$.

Hollow spherical detectors
have been recently investigated \cite{fafone} as a possible tool
for the analysis
of the relic gravitational wave background. The sensitivity of two
correlated spherical detectors could be ${\cal O}(10^{-6})$ in
$h_0^2 \Omega_{\rm GW}$ for the frequency of resonance which lies between
200 and 400 Hz. In this case the ABBN bound and the experiment will be
certainly competitive.  Dual spherical detectors \cite{cerdonio}  may reach
a sensitivity, in $h_0^2 \Omega_{\rm GW}$, which is again ${\cal O}(10^{-6})$ in the kHz
region.

Wide-band interferometers \cite{geo,tama,virgo,ligo}, a promising tool not yet 
available but close to the phase of preliminary run, will also be able to probe 
stochastic sources. 

The  observation we want to make here is very simple. Consider, for instance, the situation
where a stochastic background is detected. The bound of Eq. (\ref{bbn}) 
can help in deciding if the source is cosmological or not. 
If the background is cosmological then the bound (\ref{bbn}) will be 
satisfied. If matter--antimatter domain are present at the onset of 
BBN (a situation not impossible if hypermagnetic fields are evolving 
at the electroweak epoch) then $\Delta N_{\rm eff}$ may be rather large. 
As a consequence, provided $h_{0}^2 \Omega_{\rm GW} \leq {\cal O}(10^{-4})$, the signal 
may still be of primordial origin. A significant improvement if compared to the case 
where $\Delta N_{\rm eff} \sim {\cal O}(1)$ where $h^2 \Omega_{\rm GW} \leq {\cal O}(10^{-6})$ .

\section{Concluding remarks} 
 
CMB experiments are the present 
of experimental cosmology, GW represent a foreseeable future.
The GW spectrum ranges  over thirty decades in frequency.  
GW with (present) frequencies around
 $f_{0}\sim 10^{-18} $ Hz correspond to a 
wave-length as large as the present Hubble radius.
For these 
waves ideal detectors would be CMB experiments. 
Between few Hz and $10$ kHz is located the operating window of ground based 
interferometers. The  band of 
resonant mass detectors is around the kHz. 
Finally between few MHz and few GHz microwave 
cavities can be used as GW detectors. 

Between $10^{-18}$ Hz and $10$ kHz there are, 
roughly, 22 decades in frequency. 
The very same frequency gap, if applied to the well known electromagnetic 
spectrum,  would drive us from low-frequency radio waves up  to 
x-rays or $\gamma$-rays. As the physics explored by radio waves is very 
different from the physics probed by $\gamma$ rays 
it can be argued that the informations carried by low and high 
frequency GW must 
derive from two different physical regimes of the theory.
 
In particular, low frequency GW are sensitive to the large scale features 
of the given cosmological model and of the underlying theory of gravity, 
whereas high frequency GW are sensitive to the small scale features of a 
given cosmological model and of the underlying theory of gravity. 
For instance string theory is expected to lead to a description of gravity 
which resembles very much Einstein-Hilbert gravity at large scales 
but which can deviate from Einstein-Hilbert gravity at smaller scales. 
That is only one of the many reasons why it is very important to have 
GW detectors operating over different frequency bands.  

An apparently unrelated problem is the 
controversial origin of our magnetized Universe.
The primordial hypothesis is certainly viable. 
Furthermore, astrophysical explanations 
demand, anyway, some specific tuning whose 
origin may find explanations in cosmology.
Among other interesting signatures, stochastic 
GW backgrounds could tell us something on the nature 
and evolution of magnetic fields during the thermodynamical history
of the Universe. Few examples in this direction 
have been provided. 

\section*{Acknowledgments}
It is a pleasure to thank Eugenio Coccia and the 
comitee of the SIGRAV prize.

\end{document}